# Gas Adsorption and Diffusion Behaviors in Interfacial Systems Composed of a Polymer of Intrinsic Microporosity and Amorphous Silica: A Molecular Simulation Study

Yuta Yoshimoto,*,[†,#] Yuiko Tomita,[†,#] Kohei Sato,[†] Shiori Higashi,[‡] Masafumi Yamato,[‡] Shu Takagi,[†] Hiroyoshi Kawakami,[‡] and Ikuya Kinefuchi[†]

[†]Department of Mechanical Engineering, The University of Tokyo, 7-3-1 Hongo, Bunkyo-ku, Tokyo 113-8656, Japan

[‡]Department of Applied Chemistry, Tokyo Metropolitan University, 1-1 Minami-osawa, Hachioji, Tokyo 192-0397, Japan

# ABSTRACT

We investigate the adsorption and diffusion behaviors of $CO_2$, $CH_4$, and $N_2$ in interfacial systems composed of a polymer of intrinsic microporosity (PIM-1) and amorphous silica using grand canonical Monte Carlo (GCMC) and molecular dynamics (MD) simulations. We build model systems of mixed matrix membranes (MMMs) with PIM-1 chains sandwiched between silica surfaces. Gas adsorption analysis using GCMC simulations shows that gas molecules are preferentially adsorbed in microcavities distributed near silica surfaces, resulting in an increase in the solubility coefficients of $CO_2$, $CH_4$, and $N_2$ compared to bulk PIM-1. In contrast, diffusion coefficients obtained from MD simulations and then calibrated using the dual-mode sorption model show different tendencies depending on gas species: $CO_2$ diffusivity decreases in MMMs compared to PIM-1, whereas $CH_4$ and $N_2$ diffusivities increase. These differences are attributed to competing effects of silica surfaces: the emergence of larger pores as a result of chain packing disruption, which enhances gas diffusion, and a quadrupole–dipole interaction between gas molecules and silica surface hydroxyl groups, which retards gas diffusion. The former has a greater impact on $CH_4$ and $N_2$ diffusivities, whereas the latter has a greater impact on $CO_2$ diffusivity due to the strong quadrupole–dipole interaction between $CO_2$ and surface hydroxyls. These findings add to our understanding of gas adsorption and diffusion behaviors in the vicinity of PIM-1/silica interfaces, which are unobtainable in experimental studies.

1. INTRODUCTION

Carbon dioxide ($CO_2$) capture and separation technologies are active research areas for addressing the rapid increase in atmospheric $CO_2$ concentration causing severe environmental issues such as an acceleration of global warming,[1] sea-level rise,[2] and ocean acidification.[3] Compared to conventional yet prevalent amine-based methods,[4,5] membrane-based gas separation technologies[6,7] have several advantages: high-energy efficiency, small footprint, operational simplicity, and modularity. Specifically, separation membranes employing microporous polymers,[8–12] defined as those with pore sizes smaller than 2 nm, are promising candidates for $CO_2$ separation technologies because their pore sizes are comparable to the molecular dimensions of $CO_2$ and other smaller gases. Furthermore, microporous polymers are more advantageous than other inorganic materials in terms of low cost, film-forming ability, processability, chemical diversity, and chemical and thermal stability.

Gas transport in separation membranes follows the solution–diffusion mechanism,[13] where the permeability ($P$) is defined as a product of the diffusion coefficient ($D$) and the solubility coefficient ($S$), and the permselectivity of gas $i$ over gas $j$ is given as $P_i/P_j$. The current research efforts have been significantly devoted to surpassing the Robeson's upper bound[14] which describes the trade-off between permeability and permselectivity. To this end, various microporous polymers have been developed, including hyper-crosslinked polymers,[8] conjugated microporous polymers,[9] thermally rearranged polymers,[10] and polymers of intrinsic microporosity (PIMs).[11,12] Specifically, PIMs are a class of ladder-like polymers with rigid and contorted backbone structures. Their rigid contortion sites hinder the efficient packing of molecular chains, resulting in large free volumes, high surface areas, and unique pore structures, leading to increased $CO_2$ permeability and high permselectivity. Since the inception of archetypal PIM-1,[15,16] various PIMs and their derivatives have been developed and exhibited performances exceeding the upper bound.[11]

Mixed matrix membranes (MMMs)[17,18] formed by incorporating nonporous or porous nanofillers into polymer matrices have also attracted considerable attention as promising candidates to further push the upper bound. Incorporated nanofillers, whether porous or not, alter local polymer-chain packing and mobility, and porous nanofillers such as zeolites, zeolitic imidazolate frameworks (ZIFs), and metal-organic frameworks (MOFs) offer additional internal permeation paths. In either case, the structural properties at the polymer/nanofiller interface significantly differ from those in the bulk polymeric phase; hence gas adsorption and diffusion behaviors at the interface play a critical role in determining the separation performance in MMMs. PIM-1-based MMMs have been developed with various nanofillers such as PIM-1/silica,[19,20] PIM-1/MOFs,[21] PIM-1/ZIF-8,[22] PIM-1/multi-walled carbon nanotubes,[23] and PIM-1/porous aromatic frameworks.[24] Although improvements in gas permeability have been reported for these PIM-1-based MMMs, the molecular-level mechanisms of gas adsorption and diffusion at the polymer/nanofiller interface are yet to be fully explored. Molecular simulations are attractive techniques to probe molecular transport in interfacial systems at the atomistic scale. Some molecular simulation studies have investigated the structural properties of PIM-1/ZIF-8 and PIM-EA-TB/ZIF-8 interfaces,[25,26] gas transport through a PIM-1/ZIF-8 interface,[27] and structure and gas transport at a polyimide/zeolite interface.[28] However, to the best of our knowledge, no simulation study has been reported that investigates gas adsorption and diffusion behaviors at the PIM-1/silica interface. Although silica is a nonporous material, surface modification of silica nanoparticles has been reported to enhance their dispersibility in the polymer phase and increase the performance of PIM-1-based MMMs, surpassing the upper bound.[29] We expect a molecular-level understanding of gas adsorption and diffusion behaviors at the PIM-1/silica interface to be a valuable starting point toward a rational design of silica-incorporated MMMs.

In this study, we investigate the adsorption and diffusion behaviors of $CO_2$, $CH_4$, and $N_2$ in interfacial systems composed of PIM-1 and amorphous silica using grand canonical Monte Carlo (GCMC) and molecular dynamics (MD) simulations. We build two types of model MMMs comprising PIM-1 chains sandwiched by silica surfaces, referred to as MMM-1 and MMM-2, for systems with a single chain and two chains, respectively. Silica surfaces significantly affect PIM-1 chain packing, which is more pronounced for MMM-1 with a shorter distance between the silica surfaces. Gas adsorption analysis using GCMC simulations shows that gas molecules are preferentially adsorbed in microcavities distributed near silica surfaces, increasing the solubility coefficients of $CO_2$, $CH_4$, and $N_2$ relative to bulk PIM-1. Meanwhile, diffusion coefficients obtained from MD simulations and then calibrated using the dual-mode sorption model[30,31] show different tendencies: $CO_2$ diffusivity decreases in MMM-1 and MMM-2 compared to PIM-1, while $CH_4$ and $N_2$ diffusivities increase. These disparities are due to competing effects of silica surfaces: the emergence of larger pores resulting from chain packing disruption, which enhances gas diffusion, and a quadrupole–dipole interaction between gas molecules and silica surface hydroxyl groups, which retards gas diffusion. More specifically, the former has a greater impact on $CH_4$ and $N_2$ diffusivities, whereas the latter has a greater impact on $CO_2$ diffusivity due to $CO_2$'s strong quadrupole moment. These findings provide intriguing insights into gas adsorption and diffusion behaviors in the vicinity of PIM-1/silica interfaces, which are inaccessible to experiments.

## 2. MODEL CONSTRUCTION

### 2.1. Construction of PIM-1.

The construction of PIM-1 (Figure 1a) is performed using an *in silico* polymerization scheme as implemented in the Polymatic code.[32] This scheme allows for constructing an amorphous polymer by simulated polymerization of the constituent monomers, and its application has been proven for a range of polymers, including PIMs.[33–35] In this study, PIM-1 molecules are represented using a united-atom (UA) approach to reduce computational costs. The GAFF[36] and TraPPE-UA[37,38] force fields are used for the bonded and nonbonded interactions, where the Lorentz–Berthelot combination rules are adopted for the cross Lennard-Jones (LJ) interactions. The atomic charges are derived from quantum chemical calculations based on HF/6-31G(d) level of theory, followed by a restrained electrostatic potential (RESP) charge fitting procedure.[39] This charge scheme is the default approach in GAFF parametrization and has been shown to be suitable for a wide range of organic molecules.[36] More details on the force field parameters and atomic charges can be found elsewhere.[40] The timestep is set to be 1 fs, and the cutoff radius for the LJ and short-range electrostatic interactions is set to be 15 Å with the long-range electrostatic interactions evaluated using a particle–particle particle–mesh method[41] with a precision factor of $10^{-4}$. All MD simulations are performed using the LAMMPS software package,[42] and the visualizations of MD systems are performed with OVITO.[43]

Initially, 200 PIM-1 monomers are randomly packed in a cubic simulation box at low density (~ 0.3 g/cm$^3$) under the periodic boundary conditions. Each monomer contains predefined linker atoms that can participate in bond formation if the bonding criteria for a cutoff radius (6 Å) and participating monomer orientations are met (planarity and directionality). Energy minimization and short MD simulations are performed throughout the process to avoid high-energy conformation. This simulated polymerization is repeated until all linker atoms are used, yielding a single PIM-1 chain with a degree of polymerization (DP) of 200.

Since the PIM-1 chain thus obtained is loosely packed due to its rigid ladder structure, a 21-step MD compression/relaxation scheme[32,40] is employed to further equilibrate the PIM-1 chain (Table 1). This scheme consists of seven cycles of three MD steps: (1) NVT at a high temperature (2000 K), (2) NVT at a room temperature (300 K), and (3) NPT at a room temperature (300 K) with the pressure increased over the first three cycles (steps 1–9) and decreased for the last four cycles (steps 10–21). Here, the Nosé–Hoover style thermostat[44,45] and barostat[46,47] are used with the temperature and pressure damping parameters of 0.1 ps and 1 ps. The PIM-1 membrane thus obtained exhibits well packed structure (Figure 1b).

In the present study, we prepare five structures of PIM-1 membranes to account for statistical uncertainties. The bulk density ($\rho_{bulk}$), which is directly obtained from MD simulation, is related to experimentally obtained skeletal density ($\rho_{skel}$) as[48]

$$\frac{1}{\rho_{skel}} = \frac{1}{\rho_{bulk}} - \frac{V_{pore}}{m} \tag{1}$$

where $m$ denotes the mass of the polymer and $V_{pore}$ denotes the pore volume. The bulk density is calculated to be $\rho_{bulk} = m/V = 0.94 \pm 0.01$ g/cm$^3$ with $V$ being the cell volume. The pore volume is defined as the total amount of void space within the polymer and is computed based on the TraPPE-UA van der Waals volume of each atom of the polymer and a probe size of 0.0 Å using PoreBlazer v3.0.[49] Here, the system is discretized into cubic lattices with the side length of 0.02 Å, and the number of lattice sites accessible to the point probe is counted to evaluate the proportion of the void space. The fractional free volume is estimated to be $V_{pore}/V = 24.5\% \pm$ 0.3%, and the resultant skeletal density is $\rho_{\text{skel}} = 1.24 \pm 0.01$ g/cm$^3$, falling into the range of experimental values of 1.056–1.4 g/cm$^3$.[50,51] Additionally, the structure factors[52] of the simulated systems are consistent with experimental results[53,54] (Figure S1 in Supporting Information), ensuring faithful representation of the structural properties of PIM-1 membranes.

We also construct five PIM-1 structures with DP = 200 that are made up of eight molecular chains. The structure factor of the 8-chain membranes is almost identical to that of the 1-chain membranes, as shown in Figure S1 in Supporting Information, indicating that there are no apparent size effects. As a result, the discussion of PIM-1 membranes that follows is limited to the results obtained from the 1-chain systems.

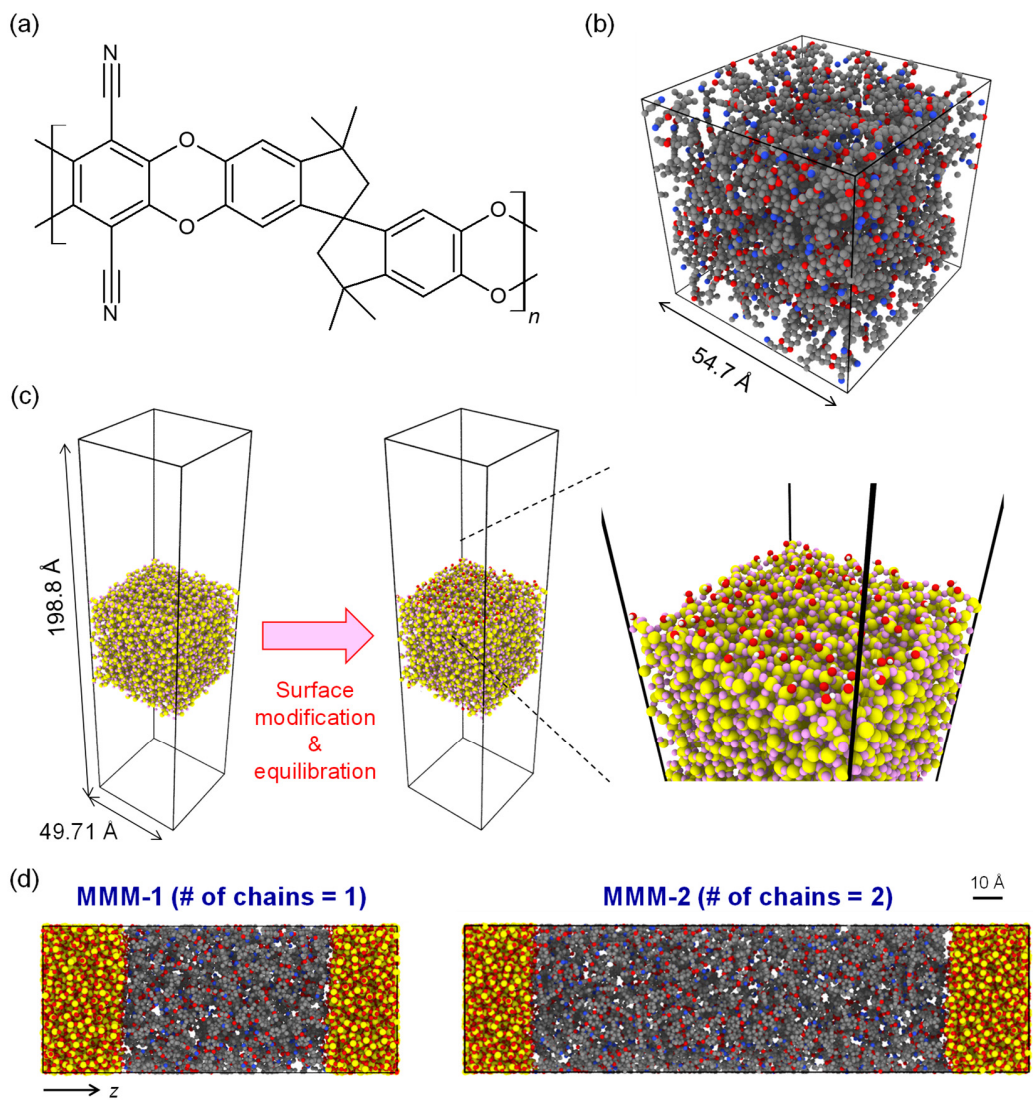

**Figure 1.** (a) PIM-1 monomer structure. (b) A representative snapshot of PIM-1 with DP = 200. Carbon, oxygen, and nitrogen atoms are represented by the gray, red, and blue spheres, respectively. (c) Fabrication of amorphous silica surfaces with a silanol group number density of 3.08 /nm$^2$. The yellow, pink, and white spheres represent silicon, bridging oxygen, and hydrogen atoms, respectively, whereas the red spheres represent hydroxide oxygens for easy identification. (d) MMMs composed of PIM-1 chains and amorphous silica. The number of PIM-1 chains is one (left) and two (right), referred to as MMM-1 and MMM-2.

**Table 1.** 21-step MD Equilibration Scheme

| Step | Ensemble | $T$ (K) | $P$ (atm) | Length (ps) |
| --- | --- | --- | --- | --- |
| 1 | NVT | 2000 | | 50 |
| 2 | NVT | 300 | | 50 |
| 3 | NPT[a]/NP$_n$T[b] | 300 | 1000 (0.02 $P_{max}$) | 50 |
| 4 | NVT | 2000 | | 50 |
| 5 | NVT | 300 | | 100 |
| 6 | NPT[a]/NP$_n$T[b] | 300 | 30000 (0.6 $P_{max}$) | 50 |
| 7 | NVT | 2000 | | 50 |
| 8 | NVT | 300 | | 100 |
| 9 | NPT[a]/NP$_n$T[b] | 300 | 50000 ($P_{max}$) | 50 |
| 10 | NVT | 2000 | | 50 |
| 11 | NVT | 300 | | 100 |
| 12 | NPT[a]/NP$_n$T[b] | 300 | 25000 (0.5 $P_{max}$) | 5 |
| 13 | NVT | 2000 | | 5 |
| 14 | NVT | 300 | | 10 |
| 15 | NPT[a]/NP$_n$T[b] | 300 | 5000 (0.1 $P_{max}$) | 5 |
| 16 | NVT | 2000 | | 5 |
| 17 | NVT | 300 | | 10 |
| 18 | NPT[a]/NP$_n$T[b] | 300 | 500 (0.01 $P_{max}$) | 5 |
| 19 | NVT | 2000 | | 5 |
| 20 | NVT | 300 | | 10 |
| 21 | NPT[a]/NP$_n$T[b] | 300 | 1 | 1600 |

[a] Construction of PIM-1 (Section 2.1). [b] Construction of PIM-1/silica systems (Section 2.3).

## 2.2. Construction of Amorphous Silica Surface.

Amorphous silica systems are constructed from β-cristobalite structure using the melt–quench method.[55] The β-cristobalite structure consists of 8232 atoms in a cubic simulation box of 50.12 × 50.12 × 50.12 Å³ under the periodic boundary conditions (Figure S2 in Supporting Information). Here, the interatomic interactions are represented by the Morse-style potential[56,57] with the cutoff radius of 10 Å, with the force field parameters provided in Table S1 in Supporting Information. Initially, the β-cristobalite structure is melted at 7000 K for 1 ns in the NVT ensemble. Next, the system is quenched to 300 K at a cooling rate of 4 K/ps, and intermediate constant-temperature simulations are performed for 100 ps at $T$ = 6000, 5000, …, 1000, and 300 K to alleviate any stresses built up in the system due to the rapid cooling.[58] The cooling rate of 4 K/ps has also been used for constructing amorphous silica at room temperature in previous studies,[55,59] leading to realistic representation of its structural properties. Finally, the system is relaxed in the NPT ensemble for 200 ps at $T$ = 300 K and $P$ = 1 atm, followed by equilibration in the NVT ensemble for 200 ps at $T$ = 300 K.

The amorphous silica structure thus obtained (Figure S2 in Supporting Information) exhibits a density of 2.23 g/cm³, which agrees with previous studies.[59,60] Moreover, partial radial distribution functions (PRDFs) of the silica structure (Figure S3 in Supporting Information) are in accordance with those obtained by Van Hoang,[59] ensuring realistic representation of the structural properties of bulk amorphous silica. The positions of the minima after the first peaks of the Si–Si, Si–O, and O–O PRDFs are $R_{Si-Si}$ = 3.46 Å, $R_{Si-O}$ = 2.03 Å, and $R_{O-O}$ = 3.03 Å, respectively, which are used as cutoff radii for defining coordination numbers ($Z_{Si-O}$ and $Z_{O-Si}$). Coordination numbers for the amorphous silica structure are provided in Tables S2 and S3 in Supporting Information.

To create silica surfaces, a vacuum layer is formed along the $z$-axis with a length sufficient to prevent interactions between top and bottom surfaces across the periodic boundary (Figure 1c). The surfaces are relaxed in the NVT ensemble for 10 ps at $T = 300$ K. Then, the dangling silicon ($Z_{Si-O} < 4$) and oxygen ($Z_{O-Si} < 2$) atoms are saturated with hydroxyl groups and hydrogens. The resultant area density of silanol groups is 3.08 /nm$^2$ (Figure 1c), which is within the range of literature values (2.6–4.6 /nm$^2$).[60] The silica surfaces contain 3% $Q^2$ (i.e., two silanol groups per superficial silicon), 53% $Q^3$ (i.e., one silanol group per superficial silicon), and 44% $Q^4$ (i.e., siloxane bridges without silanol groups) environments. This surface chemistry corresponds to neutral silanol-terminated surfaces at pH values between 2 and 4 which previously underwent thermal treatment in experiment.[61] In reality, surface chemistry varies depending on the type of substrate, surface ionization relating to pH and particle size, synthesis protocol, and prethermal treatment. More specific models reflecting these details can be considered using the INTERFACE force field,[62] which should also be investigated in future work.

To model hydroxylated silica surfaces, the force field is switched to the CHARMM water contact angle (CWCA) force field,[63] which has been designed to reproduce experimental water contact angles on silica. The force field parameters and atomic charges[55,64,65] are provided in Tables S4 and S5 in Supporting Information. We note that different proportions of the added atoms associated with the silanol groups result in a nonzero net charge. To ensure system charge neutrality, the net charge is evenly subtracted from the charges of silicon (Si) and bridging oxygen ($O_b$) atoms; the changes in their charges are less than 0.1% of the original values. Since the CWCA force field fails to capture the structural properties of bulk amorphous silica,[64] the silica framework composed of silicon and bridging oxygen atoms is kept rigid, while the surface hydroxyls are allowed to move with all O–H bond lengths maintained at their equilibrium values via the RATTLE algorithm.[66] Under these conditions, an NVT-MD simulation is performed at 300 K for 400 ps using a Langevin thermostat[67] with a damping factor of 100 fs to equilibrate the surface silanol groups.

2.3. Construction of PIM-1/Silica Hybrid Systems.

We build five MMMs from a single PIM-1 chain (DP = 200) and amorphous silica, and five MMMs from two PIM-1 chains and amorphous silica; these are referred to as MMM-1 and MMM-2, respectively (Figure 1d). In this section, we describe how to build a PIM-1/silica hybrid system for the case of MMM-1.

The PIM-1 chain obtained in Section 2.1 is first unwrapped in all directions and placed in an orthorhombic cell with the dimensions shown in Table S6 in Supporting Information. The system is then compressed and relaxed using the 21-step equilibration scheme (Table 1), resulting in cell lengths in the $x$ and $y$ directions ($L_x^{\text{PIM}} = L_y^{\text{PIM}}$) around 50 Å (Table S6 in Supporting Information), which are very close to the cell lengths in the $x$ and $y$ directions of the silica slab system ($L_x^{\text{Silica}} = L_y^{\text{Silica}} = 49.71$ Å) built in Section 2.2. Under the constant-volume condition, the PIM-1 system is slightly elongated in the $z$-direction to exactly match $L_x^{\text{PIM}}$ ($L_y^{\text{PIM}}$) with $L_x^{\text{Silica}}$ ($L_y^{\text{Silica}}$). The PIM-1 chain is then unwrapped in the $z$-direction and inserted into the vacuum layer of the silica slab system, as illustrated in Figure S4 in Supporting Information. Finally, as proposed by Semino and coworkers,[25] the constructed PIM-1/silica system is compressed and relaxed in the $z$-direction using the 21-step equilibration scheme (Table 1), yielding the equilibrated system shown in Figure 1d. To maintain the silica structure, the silica framework composed of silicon and bridging oxygen atoms is kept rigid, and compression is performed only in the $z$-direction.

In the case of MMM-2, two PIM-1 chains are first placed in an orthorhombic cell with the dimensions shown in Table S7 in Supporting Information. The following steps are the same as for MMM-1.

## 3. METHODOLOGY

### 3.1. Evaluation of Gas Solubility.

The solubility coefficient $S$ is evaluated from the slope of an adsorption isotherm in the dilute limit as

$$S = \lim_{p \to 0} \frac{C}{p} \tag{2}$$

where $C$ is the concentration of adsorbed gases and $p$ is the pressure.

In general, sorption of gases onto glassy polymers can be described using the dual-mode sorption (DMS) model[68–70] as

$$C = k_D p + \frac{C'_H b p}{1 + bp} \tag{3}$$

where $k_D$ is Henry's law constant, $C'_H$ is the Langmuir monolayer sorption capacity, and $b$ is the Langmuir affinity constant. The DMS model is based on two types of sorption sites subject to Henry's law dissolution and Langmuir-type sorption and has been shown to accurately represent the gas sorption in PIM-1.[71,72]

This study evaluates the adsorption of $CO_2$, $CH_4$, and $N_2$ in PIM-1, MMM-1, and MMM-2 through grand canonical Monte Carlo (GCMC) simulations using the LAMMPS software package.[42] GCMC simulations perform exchanges of molecules with an imaginary gas reservoir at the specified temperature ($T$) and chemical potential ($\mu$).[73] The gas molecules are modeled with the TraPPE-UA force field,[37,38] where $CO_2$ and $N_2$ are treated as rigid molecules, and $CH_4$ is represented by the united-atom model with a single interaction site. GCMC simulations for a variety of chemical potentials are performed to obtain adsorption isotherms for each gas species in each membrane. GCMC simulations are run at 300 K in the $\mu$VT ensemble for each chemical potential (or pressure), with the membrane framework held rigid during gas adsorption. After each GCMC step, a 100-ps MD simulation is performed at 300 K in the NVT ensemble to equilibrate the system. This GCMC–MD cycle is repeated until the number of gas molecules in the system converges sufficiently. Before the gas adsorption simulations in the membranes, the relationship between the chemical potential and pressure is obtained for each gas species through bulk gas-phase GCMC simulations to take account of the nonideality of the gases at high pressures (Figure S5 in Supporting Information).

We note that the present study does not consider plasticization and swelling of the membranes upon gas adsorption. It has been reported that gas adsorption at high pressures induces the swelling of PIM-1,[74–76] increasing the free volume and chain segment mobility. Although some studies have proposed the methods for simulating the swelling of polymeric membranes,[77–79] rigid nature of the silica framework in the present study allows for the volume dilation only in the interface-normal direction (Figure 1d), prohibiting a proper evaluation of the membrane swelling behavior. Rather, the present study mainly focuses on the PIM-1/silica interface that leads to the disruption of polymer-chain packing and generation of microcavities in the vicinity of the interface; these factors significantly affect adsorption and diffusion behaviors of the gases in the membranes.

## 3.2. Evaluation of Gas Diffusivity.

The self-diffusion coefficient $D$ can be evaluated from MD simulation using the Einstein relationship[52] as

$$D = \frac{1}{2Nd} \lim_{t \to \infty} \frac{1}{t} \langle \sum_{i=1}^{N} |\mathbf{r}_i(t) - \mathbf{r}_i(0)|^2 \rangle \qquad (4)$$

where $N$ is the number of gas molecules, $d$ is the dimensionality of the diffusing molecules under consideration, and $\mathbf{r}_i(t)$ is the center-of-mass position of molecule $i$ at time $t$. The term in brackets is the so-called mean-squared displacement (MSD).

To determine the self-diffusion coefficients of $CO_2$, $CH_4$, and $N_2$ in PIM-1, MMM-1, and MMM-2 at dilute conditions, 10 gas molecules are randomly inserted into the pore space within the polymer region for each gas–membrane combination. The gas molecules' trajectories are then sampled in the NVE ensemble, while the temperatures of the membrane frameworks are kept at 300 K using the Nosé–Hoover thermostat.[44,45] The three-dimensional MSD of the gas molecules ($d = 3$) is evaluated in PIM-1, while the two-dimensional MSDs in the $xy$-plane ($d = 2$) are calculated in MMM-1 and MMM-2 because the PIM-1 regions are bounded by the nonporous silica frameworks (Figure 1d). We note that the self-, collective, and transport diffusion coefficients coincide in the limit of zero loading of gas molecules.[80,81] In the present study, inserting 10 gas molecules into each membrane corresponds to the gas pressures of at most 0.02, 0.2, and 0.6 atm for $CO_2$, $CH_4$, and $N_2$, respectively (see Figure 4). For these small pressure values, the obtained self-diffusion coefficients would be good proxies for the transport diffusion coefficients. To confirm convergence to the zero-loading diffusion coefficient, it is necessary to calculate self- and transport diffusion coefficients using equilibrium and nonequilibrium MD simulations with varied gas loadings[82] and extrapolate the diffusivity–loading curves to the zero-loading limit, which is beyond the scope of this study.

Unfortunately, diffusion coefficients calculated using MD simulations cannot be directly compared to those obtained using time-lag experiments. Namely, the diffusion coefficient determined by transient permeation experiments is given as[83,84]

$$D_\theta = \frac{l^2}{6\theta} \tag{5}$$

where $l$ is the thickness of a polymer membrane, and $\theta$ is the diffusion time-lag. Based on the DMS model,[68–70] the diffusion time-lag of gases in a glassy polymer can be related to the diffusion coefficient associated with Henry's law dissolution ($D_D$) using the DMS fitting parameters (eq 3) as[30,31,85]

$$D_D = \frac{l^2}{6\theta} f(F, K, \eta) \tag{6}$$

$$F = \frac{D_H}{D_D} \tag{7}$$

$$K = \frac{C'_H b}{k_D} \tag{8}$$

$$\eta = bp_2 \tag{9}$$

where $D_H$ denotes the diffusion coefficient associated with the Langmuir-type sorption and $p_2$ is the gas pressure upstream of the polymer film. The functional form of $f$ is given by[31]

$$f(F, K, \eta) = \frac{1 + K[f_0 + FKf_1 + (FK)^2 f_2] + FKf_3 + (FK)^2 f_4}{\left(1 + \frac{FK}{1+\eta}\right)^3} \quad (10)$$

where

$$f_0 = \frac{6}{\eta^3}\left[\frac{\eta^2}{2} + \eta - (1+\eta)\ln(1+\eta)\right]$$

$$f_1 = \frac{6}{\eta^3}\left[\frac{\eta}{2} - \frac{3\eta}{2(1+\eta)} + \frac{\ln(1+\eta)}{1+\eta}\right]$$

$$f_2 = \frac{6}{\eta^3}\left[\frac{1}{6} - \frac{1}{2(1+\eta)} + \frac{1}{2(1+\eta)^2} - \frac{1}{6(1+\eta)^3}\right] \quad (11)$$

$$f_3 = \frac{6}{\eta^3}\left[-\frac{3}{2}\eta + \frac{\eta}{2(1+\eta)} + (1+\eta)\ln(1+\eta)\right]$$

$$f_4 = \frac{6}{\eta^3}\left[\frac{1}{2} - \frac{1}{2(1+\eta)^2} - \frac{\ln(1+\eta)}{1+\eta}\right]$$

This model considers the partial immobilization of gases held by the Langmuir mode in a glassy polymer; $F = 0$ and 1 correspond to total immobilization and no immobilization, respectively, while partial immobilization is represented by $0 < F < 1$.

Because diffusion coefficients obtained from MD simulations are proxies for $D_D$ rather than $D_\theta$, we estimate $D_\theta$ by dividing the calculated diffusion coefficients by $f(F, K, \eta)$ according to eqs 5 and 6. This calibration offers a more reasonable comparison with the diffusion coefficients obtained from time-lag experiments. The DMS fitting parameters ($k_D$, $C'_H$, and $b$) can be evaluated from adsorption isotherms obtained by GCMC simulations as described in Section 3.1, whereas $F$ and $p_2$ are adjustable parameters. Here, $p_2$ is assumed to be 1 atm, a common choice for time-lag experiments. Meanwhile, the determination of $F$ is nontrivial due to the difficulty of calculating $D_H$. In this study, we adopt $F = 0.026$, $0.009$, and $0.026$ for $CO_2$, $CH_4$, and $N_2$, respectively; these values were estimated by Robeson and coworkers[85] using permeation data of PIM-1.[71]

## 4. RESULTS & DISCUSSION

### 4.1. Structural Characteristics of the Membranes.

Figure 2 depicts the $z$-direction density distributions of representative MMM-1 and MMM-2. Here, we define three regions in MMM-1 and MMM-2: a PIM-1 region, a silica region, and a PIM-1/silica mixing region. The PIM-1 region is defined as the presence of only PIM-1, whereas the silica region is defined as the presence of only silica. The PIM-1/silica mixing region is defined as having nonzero densities of both PIM-1 and silica, as indicated by the blue dashed lines in Figure 2. The thicknesses of the PIM-1/silica mixing regions are in the range of 4–5 Å. We note that the average density of the PIM-1 region in MMM-1 is approximately 0.85 g/cm$^3$, which is less than the density of bulk PIM-1 (0.94 ± 0.01 g/cm$^3$) indicated by the gray line in Figure 2a. This decrease in the density of the PIM-1 region originates from the disruption of PIM-1 chain packing caused by the presence of silica surfaces. A decrease in polymer-phase density in the presence of inorganic fillers has also been reported for PIM-1/MOF composites.[25] In contrast, the average density of the PIM-1 region in MMM-2 (Figure 2b) is around 0.91 g/cm$^3$, which is closer to the density of bulk PIM-1 (0.94 ± 0.01 g/cm$^3$). This result indicates that the effect of the silica surfaces on chain packing is reduced for MMM-2 with a greater distance between the silica surfaces than for MMM-1.

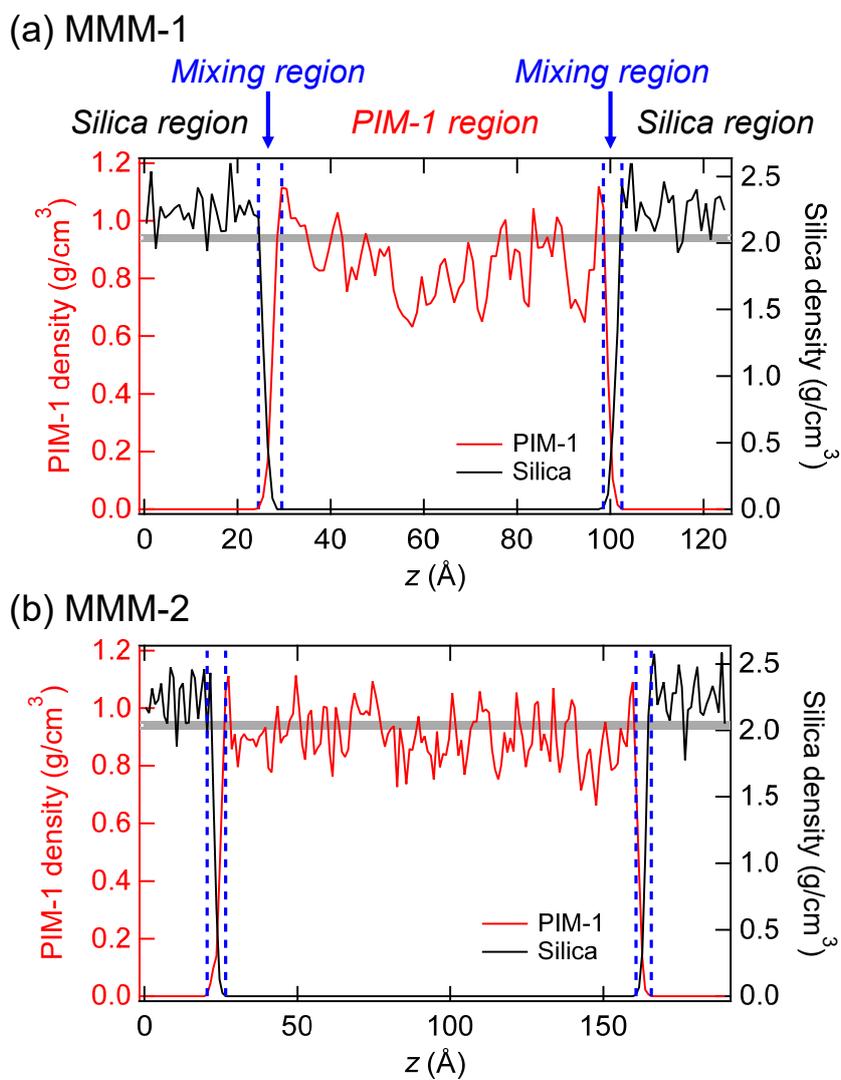

**Figure 2.** Density distributions in the *z*-direction of representative MMM-1 (a) and MMM-2 (b). The gray lines represent the bulk PIM-1 average density (0.94 ± 0.01 g/cm$^3$). The blue dashed lines represent PIM-1/silica mixing regions, which have nonzero densities of both PIM-1 and silica.

To further characterize free volumes in the membranes, we evaluate geometric pore size distributions (PSDs) in representative PIM-1 and polymer regions of representative MMM-1 and MMM-2 (i.e., PIM-1 and PIM-1/silica mixing regions), as shown in Figure 3a. The PSD is defined here as a statistical distribution of the diameter of the largest sphere that can be fitted inside a pore at a given point,[86,87] and is calculated using PoreBlazer v3.0[49] based on the van der Waals volume of each atom of PIM-1 and silica. The PSD of PIM-1 has a peak around 6–7 Å, whereas the PSD of MMM-1 is broader, exhibiting larger pores with diameters > 10 Å. Figure 3b depicts the PSD in the polymer region of MMM-1 in three dimensions, highlighting the larger pores within the polymer phase. The formation of larger pores in MMM-1 is caused by the disruption of chain packing caused by the silica surfaces, which results in a decrease in polymer-phase density, as shown in Figure 2a. In contrast, Figure 3a shows that the PSD of MMM-2 is narrower than that of MMM-1 and has a similar distribution to that of PIM-1. This result indicates that the greater distance between the silica surfaces in MMM-2 results in a more packed structure in the PIM-1 region than in MMM-1. Additionally, there exist small cavities with the diameters $\lesssim$ 3 Å in MMM-1 and MMM-2 compared to PIM-1, as shown in the inset in Figure 3a. These microcavities are mostly found in the PIM-1/silica mixing regions (Figure 3c), contributing to the increased gas solubility in MMM-1 and MMM-2 over PIM-1, as discussed in Section 4.2.

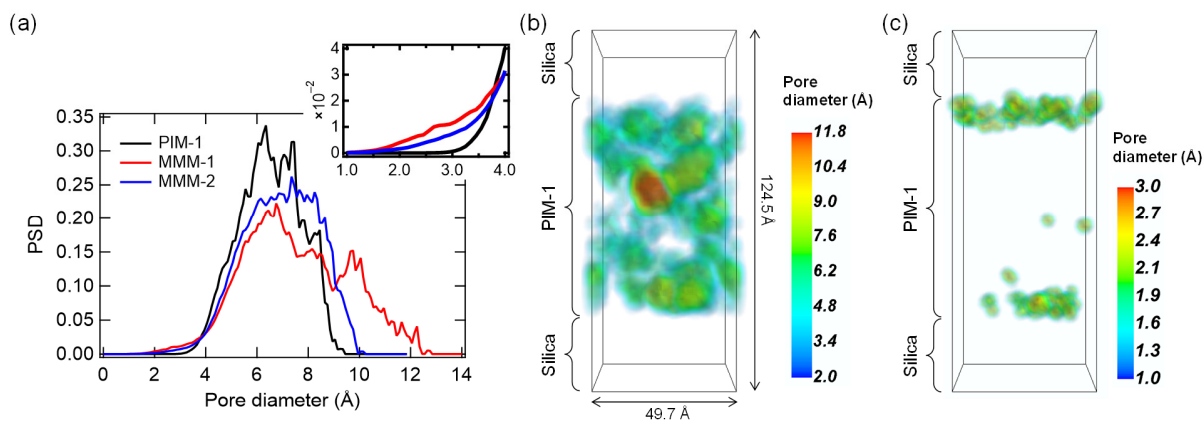

**Figure 3.** (a) Pore size distributions (PSDs) in representative PIM-1 and polymer regions of representative MMM-1 and MMM-2 (i.e., PIM-1 and PIM-1/silica mixing regions). The inset shows a zoomed-in view of the PSDs with pore diameters ranging from 1 to 4 Å. (b) A three-dimensional visualization of the PSD in MMM-1's polymer region, with the PIM-1 and silica omitted for clarity. Mayavi[88] is used for the visualization. (c) A visualization of the microcavities (≤ 3 Å) distributed in the PIM-1/silica mixing regions of MMM-1.

### 4.2. Gas Adsorption Behavior.

Figure 4a depicts the adsorption isotherms of $CO_2$, $CH_4$, and $N_2$ in PIM-1. All of the isotherms have convex pressure curves, and $CO_2$ is the most condensable, with an initial loading at low pressures that is significantly higher than those of $CH_4$ and $N_2$. The $CO_2$, $CH_4$, and $N_2$ solubility coefficients in PIM-1 are calculated using eq 2 and tabulated in Table 2 compared to experimentally obtained results. Despite some scatter in the experimental data, the computed solubility coefficients agree well with the experimental values, indicating that our model systems successfully represent gas adsorption behavior. Admittedly, the $CO_2$ and $N_2$ solubility coefficients in PIM-1 are slightly overestimated compared to the experimental values, which may be due to the inadequacy of the current force field. Nevertheless, considering their slight differences, our model systems can be safely used to quantitatively predict the solubility coefficients.

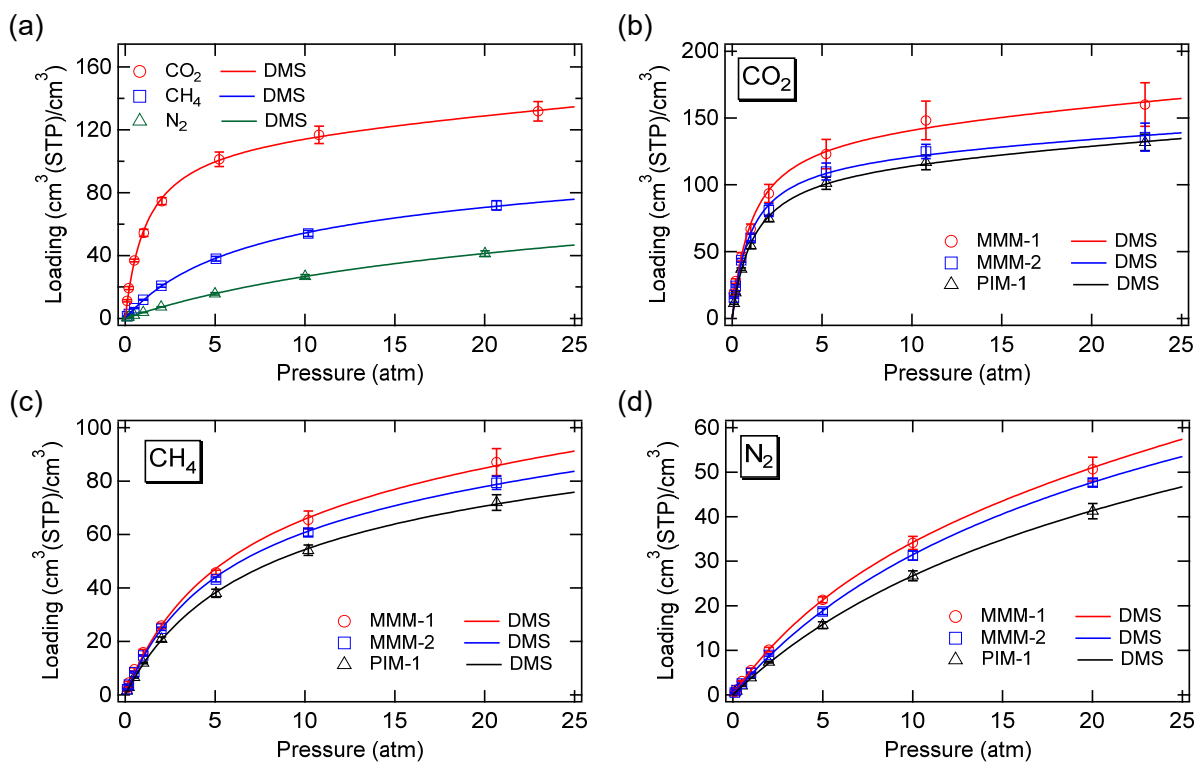

**Figure 4.** (a) $CO_2$, $CH_4$, and $N_2$ adsorption isotherms in PIM-1. (b–d) Adsorption isotherms of $CO_2$ (b), $CH_4$ (c), and $N_2$ (d) in PIM-1, MMM-1, and MMM-2. The DMS model (eq 3) is used to fit the gas loadings using the parameters listed in Table 3.

**Table 2.** Solubility Coefficients [×10$^{-2}$ cm$^3$(STP)/cm$^3$·cmHg] of CO$_2$, CH$_4$, and N$_2$ in PIM-1, MMM-1, and MMM-2

| Membrane | Gas | $S_{sim}$[a] | $S_{exp}$[b] |
|---|---|---|---|
| PIM-1 | CO$_2$ | 102.9 ± 1.4 | 29.6,[89] 88.0[90] |
| | CH$_4$ | 17.4 ± 0.3 | 9.37,[89] 18.0[90] |
| | N$_2$ | 5.43 ± 0.06 | 2.05,[89] 4.2[90] |
| MMM-1 | CO$_2$ | 134.0 ± 3.8 | — |
| | CH$_4$ | 26.1 ± 0.7 | — |
| | N$_2$ | 8.16 ± 0.21 | — |
| MMM-2 | CO$_2$ | 123.8 ± 1.5 | — |
| | CH$_4$ | 22.9 ± 0.3 | — |
| | N$_2$ | 6.79 ± 0.13 | — |

[a] Each uncertainty represents a standard error obtained from five independent simulations. [b] The experimental temperature and pressure are 303 K and 0.2 bar,[90] and 308 K and 4 atm.[89]

Figure 4b depicts CO$_2$ adsorption isotherms in PIM-1, MMM-1, and MMM-2, with gas loadings in MMM-1 and MMM-2 calculated using the amount of CO$_2$ adsorbed within the polymer regions (i.e., PIM-1 and PIM-1/silica mixing regions). MMM-1 has the highest CO$_2$ loading, followed by MMM-2 and PIM-1, and the same trends can be seen in the CH$_4$ and N$_2$ isotherms, as shown in Figures 4c and 4d. MMM-1 has the highest solubility coefficient for each gas species, followed by MMM-2 and PIM-1, as shown in Table 2. To explain the increases in solubility coefficients in MMM-1 and MMM-2, gas density distributions in representative MMM-1 are evaluated for a range of gas pressures, as shown in Figure 5, and those in representative MMM-2 are shown in Figure S6 in Supporting Information. As shown in Figure 5a, for $CO_2$ pressure < 1 atm, more $CO_2$ molecules are adsorbed in the PIM-1/silica mixing regions (defined in Figure 2a) than in the PIM-1 region. This preferential adsorption of $CO_2$ in the PIM-1/silica mixing regions at low pressure is attributed to the presence of energetically favorable adsorption sites, such as the microcavities shown in Figure 3c, contributing to an increase in the solubility coefficient in MMM-1. The amount of $CO_2$ adsorbed in the PIM-1 region increases as $CO_2$ pressure increases, resulting in less pronounced effects of the PIM-1/silica mixing regions. As shown in Figures 5b and 5c, these tendencies also hold true for $CH_4$ and $N_2$ adsorptions in MMM-1. Because the proportion of PIM-1/silica mixing regions to a PIM-1 region in MMM-2 is smaller than in MMM-1, the enhancement of solubility coefficients is less pronounced in MMM-2.

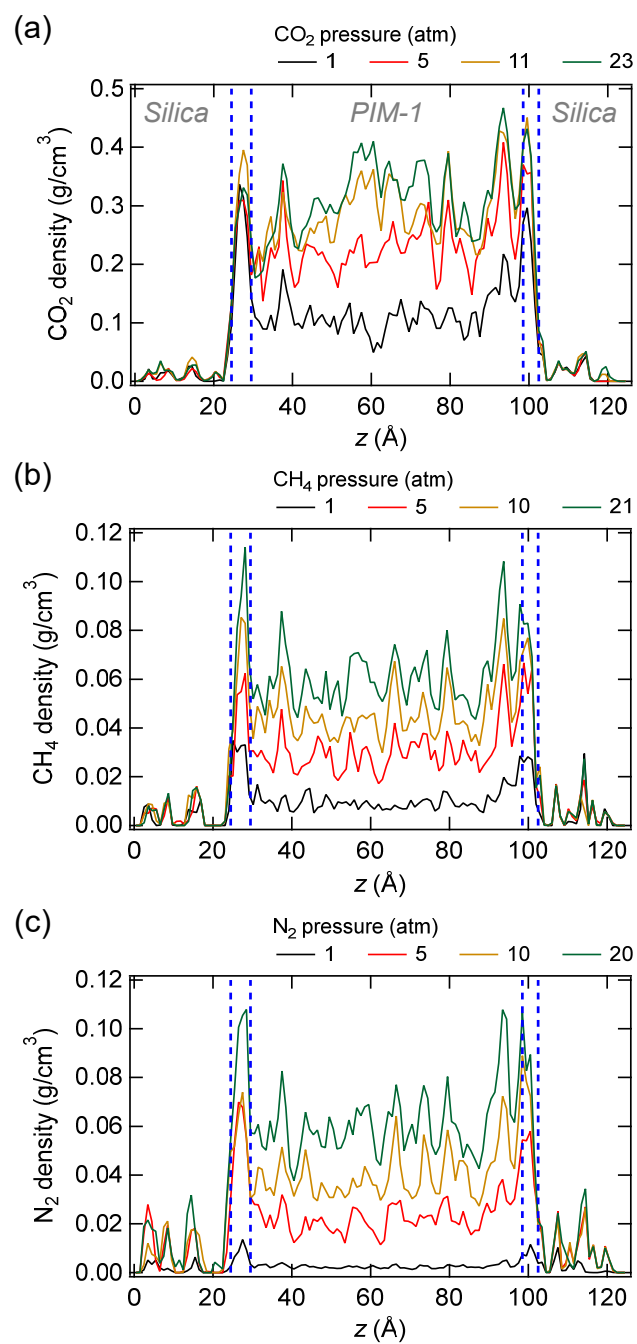

**Figure 5.** Gas density distributions in representative MMM-1; $CO_2$ at pressures ranging from 1 to 23 atm (a), $CH_4$ at pressures ranging from 1 to 21 atm (b), and $N_2$ at pressures ranging from 1 to 20 atm (c). According to Figure 2a, the blue dashed lines represent PIM-1/silica mixing regions.

Figure 6 shows solubility *versus* solubility selectivity for the gas pairs of $CO_2/CH_4$ and $CO_2/N_2$. While the $CO_2$ solubility increases in MMM-1 and MMM-2 compared to PIM-1, the solubility selectivities for $CO_2/CH_4$ and $CO_2/N_2$ decrease only slightly, exhibiting similar tendencies to those observed in thermo-oxidatively crosslinked PIM-1 films incorporated with silica nanoparticles ($CO_2/CH_4$ selectivity of 3.7–4.6 and $CO_2/N_2$ selectivity of 11.2–14.6).[20] We note that direct comparison of the calculated results with experimentally obtained findings[19,20] is unfeasible because silica nanoparticles are dispersed in a polymer matrix with weight ratios up to ~ 40 wt% in realistic MMMs, whereas our model systems of MMM-1 and MMM-2 are considered to represent PIM-1/silica interfacial systems. Nonetheless, our model systems capture a moderate dependence of the solubility selectivity on the proportion of the silica region, implying that the gas adsorption behavior at the PIM-1/silica interfaces is appropriately represented.

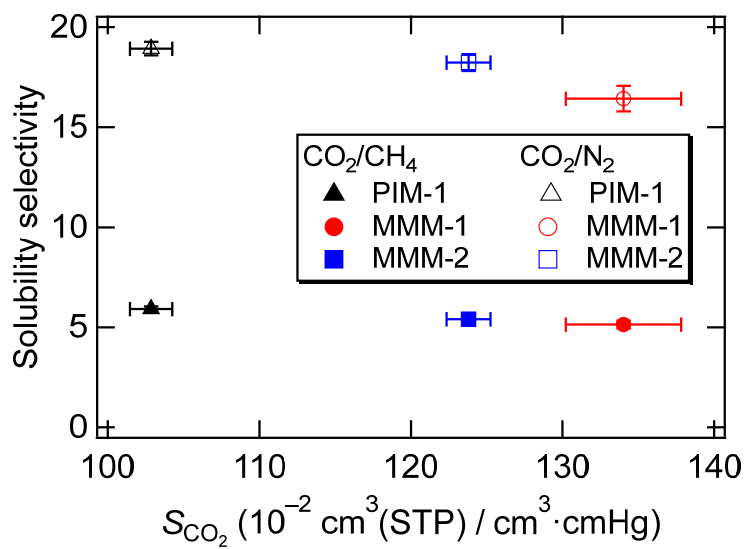

**Figure 6.** Solubility *versus* solubility selectivity for the gas pairs $CO_2/CH_4$ and $CO_2/N_2$. Each $CO_2$ solubility error bar represents a standard error derived from five independent simulations, whereas each uncertainty of solubility selectivity is estimated by error propagation using the uncertainties of respective solubilities.

Finally, as shown in Figure 4, the adsorption isotherms of $CO_2$, $CH_4$, and $N_2$ in PIM-1, MMM-1, and MMM-2 are fitted by the DMS model (eq 3) with the fitting parameters provided in Table 3. Both Henry's law constant ($k_D$) and Langmuir monolayer sorption capacity ($C'_H$) tend to increase for highly condensable gases ($CO_2 > CH_4 > N_2$). We note that the DMS model is highly sensitive to the original pressure range and extrapolated distance,[91] and a detailed analysis of the DMS fitting parameters is beyond the scope of this study. However, as described in Section 3.2, the DMS fitting parameters can be used for calibrating diffusion coefficients obtained from MD simulations (Section 4.3).

**Table 3.** DMS Fitting Parameters for $CO_2$, $CH_4$, and $N_2$ Adsorbed in PIM-1, MMM-1, and MMM-2

| Membrane | Gas | $k_D$ (cm³(STP)/cm³·atm) | $C'_H$ (cm³(STP)/cm³) | $b$ (atm⁻¹) |
|---|---|---|---|---|
| PIM-1 | $CO_2$ | 0.906 | 117.2 | 0.861 |
| | $CH_4$ | 0.401 | 82.83 | 0.155 |
| | $N_2$ | 0.328 | 67.47 | 0.0532 |
| MMM-1 | $CO_2$ | 1.05 | 144.6 | 0.906 |
| | $CH_4$ | 0.614 | 92.74 | 0.180 |
| | $N_2$ | 0.717 | 56.93 | 0.0905 |
| MMM-2 | $CO_2$ | 0.764 | 124.7 | 0.989 |
| | $CH_4$ | 0.579 | 83.57 | 0.193 |
| | $N_2$ | 0.307 | 77.59 | 0.0577 |

4.3. Gas Diffusion Behavior.

Figure 7 shows the MSDs of $CO_2$, $CH_4$, and $N_2$ in representative PIM-1 (a), MMM-1 (b), and MMM-2 (c). As mentioned in Section 3.2, three-dimensional MSDs are calculated for the gases in PIM-1 (a), whereas two-dimensional MSDs in the $xy$-plane are calculated in MMM-1 (b) and MMM-2 (c). The gas molecules initially undergo short-time ballistic diffusion with the MSDs proportional to $t^2$, followed by the so-called anomalous diffusion[92] characterized by the MSDs proportional to $t^n$ $(0 < n < 1)$. It takes at least 10 ns for the gas molecules to finally arrive at the normal-diffusion regime with the MSD proportional to $t$ for each gas–membrane combination. The emergence of the long-time anomalous diffusion regime is due to the highly inhomogeneous nature of the interconnecting pore networks within the polymer phase, and similar trends have been observed for the gas diffusion in amorphous polyisobutylene[93] and fluorinated polyimides.[94] In this study, long-time trajectories (> 500 ns) of the gas molecules are obtained from MD simulations, from which 50-ns trajectories for PIM-1 and 20-ns trajectories for MMM-1 and MMM-2 are extracted by shifting the time origin to accurately estimate the self-diffusion coefficients in the normal-diffusion regime.

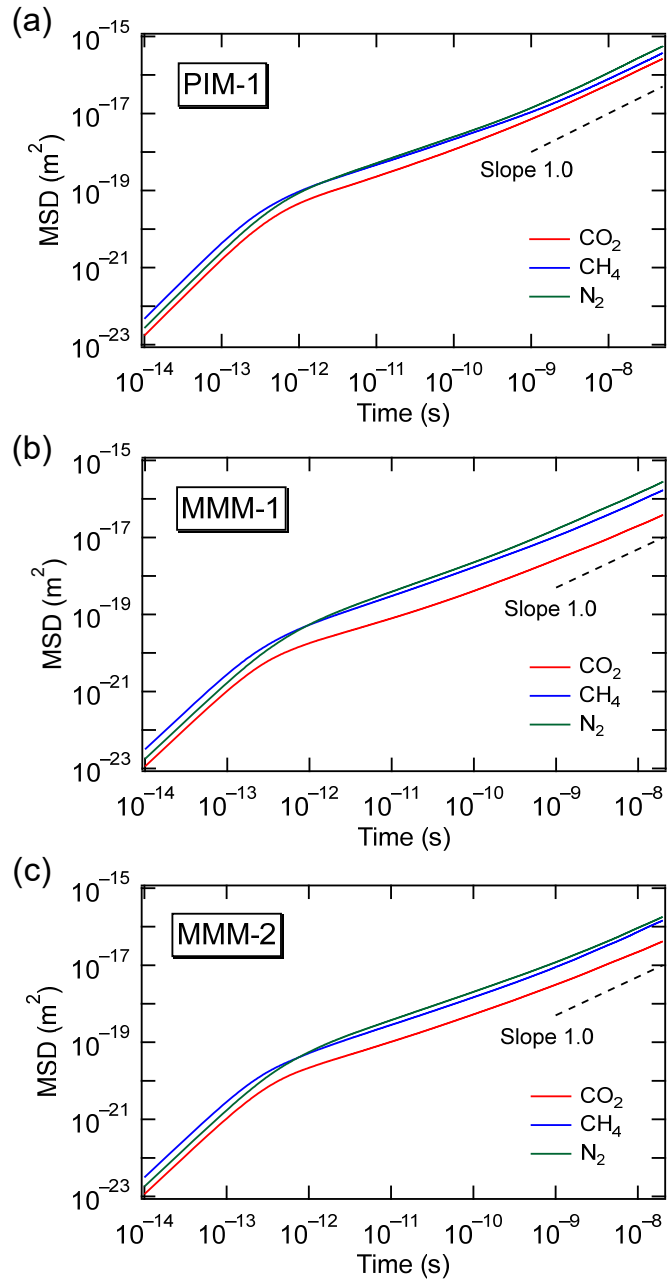

**Figure 7.** Mean-squared displacements (MSDs) of $CO_2$, $CH_4$, and $N_2$ in representative PIM-1 (a), MMM-1 (b), and MMM-2 (c). Three-dimensional MSDs are calculated for the gases in PIM-1 (a), whereas two-dimensional MSDs in the *xy*-plane are calculated in MMM-1 (b) and MMM-2 (c). The dashed lines with slopes of unity are plotted for the identification of normal-diffusion regimes.

Table 4 shows the self-diffusion coefficients of $CO_2$, $CH_4$, and $N_2$ in PIM-1 obtained from the slopes of the MSDs ($D_{\text{sim}}$). $N_2$ has the highest self-diffusion coefficient [(1680 ± 348) × $10^{-8}$ cm$^2$/s] followed by $CH_4$ [(1062 ± 241) × $10^{-8}$ cm$^2$/s], and $CO_2$ [(791 ± 112) × $10^{-8}$ cm$^2$/s]. The uncertainty of each diffusion coefficient representing a standard error obtained from five independent simulations is significant compared to the mean value, highlighting the highly inhomogeneous nature of the pore network in the respective membranes. As mentioned in Section 3.2, because it is not feasible to directly compare the diffusion coefficients obtained from MD simulations with experimental results obtained from transient permeation experiments, the calibrated diffusion coefficients $D_{\text{sim}}^{\text{calib}}$ are obtained by dividing $D_{\text{sim}}$ with $f(F, K, \eta)$ using the DMS fitting parameters in Table 3 according to eqs 5 and 6. As shown in Table 4, the calibrated diffusion coefficients of $CO_2$, $CH_4$, and $N_2$ in PIM-1 are (29.9 ± 4.2) × $10^{-8}$ cm$^2$/s, (43.8 ± 9.9) × $10^{-8}$ cm$^2$/s, and (184.1 ± 38.1) × $10^{-8}$ cm$^2$/s, respectively. Considering the significant scatter in the experimental values, the calibrated diffusion coefficients show reasonable agreement with the experimental values; however, they do not follow the experimental trend[89,90] that $CO_2$ diffusivity is higher than or comparable to $N_2$ diffusivity with $CH_4$ diffusivity being the lowest. Some factors are responsible for these discrepancies. (1) When comparing calibrated diffusion coefficients to experimental values, concentration-independent diffusivity is assumed; however, it does not necessarily hold depending on the gas species.[81] (2) The DMS fitting parameters are highly sensitive to the pressure range,[91] and the values of $F$ (eq 7) have significant uncertainties.[85] (3) The TraPPE-UA force field[37,38] used for the nonbonded interaction has been optimized for the quantitative prediction of thermophysical properties, particularly phase equilibria, over a wide range of physical conditions; however this does not guarantee the quantitative prediction of transport coefficients. (4) Although the *static* structure factor of the simulated PIM-1 are consistent with experimental data (Figure S1 in Supporting Information), a *dynamic* chain mobility in the simulated PIM-1 may be different from that in experimental PIM-1 due to a short-chain length of the former (DP = 200) compared to a typical value of the latter (DP > 500);[16,95] this may result in differences of gas diffusion behavior.

**Table 4.** Diffusion Coefficients [×10$^{-8}$ cm$^2$/s] of $CO_2$, $CH_4$, and $N_2$ in PIM-1, MMM-1, and MMM-2

| Membrane | Gas | $D_{\text{sim}}$[a] | $D_{\text{sim}}^{\text{calib}}$[b] | $D_{\text{exp}}$[c] |
|---|---|---|---|---|
| PIM-1 | $CO_2$ | 791 ± 112 | 29.9 ± 4.2 | 26,[90] 120[89] |
| | $CH_4$ | 1062 ± 241 | 43.8 ± 9.9 | 6.8,[90] 40[89] |
| | $N_2$ | 1680 ± 348 | 184.1 ± 38.1 | 22,[90] 120[89] |
| MMM-1 | $CO_2$ | 485 ± 78 | 17.7 ± 2.9 | — |
| | $CH_4$ | 1857 ± 495 | 87.4 ± 23.3 | — |
| | $N_2$ | 3563 ± 621 | 532.8 ± 92.9 | — |
| MMM-2 | $CO_2$ | 479 ± 45 | 16.5 ± 1.5 | — |
| | $CH_4$ | 1690 ± 160 | 78.6 ± 7.4 | — |
| | $N_2$ | 2158 ± 145 | 194.5 ± 13.1 | — |

[a] $D_{\text{sim}}$ refers to the diffusion coefficient obtained from MD simulations via eq 4. Each uncertainty represents a standard error obtained from five independent simulations. [b] $D_{\text{sim}}^{\text{calib}}$ refers to the diffusion coefficient calibrated to approximate $D_\theta$ in eq 5. [c] The experimental temperature and pressure are 303 K and 0.2 bar,[90] and 308 K and 4 atm.[89]

Despite these potential drawbacks, our model systems do provide some qualitative insights into the gas diffusion behavior in the vicinity of PIM-1/silica interfaces. As shown in Table 4,

MMM-1 has the highest $N_2$ diffusivity [$(532.8 \pm 92.9) \times 10^{-8}$ cm$^2$/s], followed by MMM-2 [$(194.5 \pm 13.1) \times 10^{-8}$ cm$^2$/s], and PIM-1 [$(184.1 \pm 38.1) \times 10^{-8}$ cm$^2$/s], and the same trend is observed for the $CH_4$ diffusivity. The increased $N_2$ and $CH_4$ diffusivity in MMM-1 and MMM-2 is due to larger pores in the polymer phase (Figure 3), which are caused by the disruption of chain packing caused by the presence of silica surfaces. Because MMM-2 has a greater distance between the silica surfaces than MMM-1, the degree of chain packing disruption is reduced, and the $N_2$ ($CH_4$) diffusivity in MMM-2 is closer to that in PIM-1. Surprisingly, $CO_2$ diffusivity exhibits a different tendency: the diffusion coefficient of $CO_2$ is $(17.7 \pm 2.9) \times 10^{-8}$ cm$^2$/s in MMM-1 and $(16.5 \pm 1.5) \times 10^{-8}$ cm$^2$/s in MMM-2, both of which are lower than the diffusion coefficient of $CO_2$ in PIM-1 [$(29.9 \pm 4.2) \times 10^{-8}$ cm$^2$/s]. The reduction in $CO_2$ diffusivity in MMM-1 and MMM-2 is due to a strong quadrupole–dipole interaction between $CO_2$ and the hydroxyl groups of the silica surfaces, which retards $CO_2$ diffusion, whereas $N_2$ has a weaker quadrupole moment and $CH_4$ does not. It has been reported that incorporating polar hydroxyl groups into polyimides with intrinsic microporosity improves $CO_2$ adsorption due to their strong dipole moments.[96] The strong quadrupole–dipole interaction between $CO_2$ and surface hydroxyls is also shown in Figure 5, where a significant amount of $CO_2$ is adsorbed even at low pressure in the PIM-1/silica mixing regions. Here, we see two competing effects of silica surfaces: the formation of larger pores, which improves gas diffusion, and the quadrupole–dipole interaction, which retards gas diffusion. More specifically, our findings show that the former has a greater impact on $N_2$ and $CH_4$ diffusivity, whereas the latter has a greater impact on $CO_2$, resulting in a reduction in $CO_2$ diffusivity.

Experimentally, the addition of silica nanoparticles into a PIM-1 matrix leads to an increase in $CO_2$ diffusivity.[20] The difference between our simulation and experimental results is because aggregated silica nanoparticles are dispersed in a polymer matrix with the weight ratio of up to ~ 40 wt% in experiments,[19,20] whereas our MMM models highlight PIM-1/silica interfacial systems. Nonetheless, our findings suggest that the presence of silica surfaces results in different diffusion behaviors of gases in the vicinity of PIM-1/silica interfaces, providing intriguing molecular insights that are unobtainable by experiments. Finally, we opine that to model more realistic MMMs considering the dispersion of aggregated nanoparticles in a PIM-1 matrix, a coarse-grained (CG) MD technique would be an attractive choice because CGMD allows for simulation at a much larger spatiotemporal scale than can be achieved with all-atom MD.[97–99] However, the CG force field requires precise parametrization to reproduce static/dynamic properties of interest.

## 5. CONCLUSIONS

In this study, we investigate the adsorption and diffusion behaviors of $CO_2$, $CH_4$, and $N_2$ in interfacial systems composed of a polymer of intrinsic microporosity (PIM-1) and amorphous silica using grand canonical Monte Carlo (GCMC) and molecular dynamics (MD) simulations. PIM-1 chains with a degree of polymerization of 200 are sandwiched between silica surfaces in our model systems of mixed matrix membranes (MMMs), which are built using an MD compression/relaxation scheme. When compared to bulk PIM-1, the MMM with a single PIM-1 chain (MMM-1) has larger pores with diameters > 10 Å within the polymer phase, which are caused by chain packing disruption due to the presence of silica surfaces. This effect is mitigated in the MMM having two PIM-1 chains (MMM-2) with a greater distance between the silica surfaces than MMM-1. In addition, microcavities with diameters ≲ 3 Å are formed near the silica surfaces in MMM-1 and MMM-2. Gas adsorption analysis using GCMC simulations shows that gas molecules are preferentially adsorbed in these microcavities, resulting in an increase in solubility coefficients in MMM-1 and MMM-2, with only minor decreases in solubility selectivities ($CO_2/CH_4$ and $CO_2/N_2$). Because the proportion of PIM-1/silica mixing regions with microcavities is smaller in MMM-2 than in MMM-1, the solubility enhancement is less pronounced. Furthermore, mean-squared displacements (MSDs) of $CO_2$, $CH_4$, and $N_2$ obtained from MD simulations show that it takes at least 10 ns for the gas molecules to reach a normal-diffusion regime, emphasizing the highly inhomogeneous nature of pore networks within the membranes. The diffusion coefficients calculated from the slopes of the MSDs are calibrated using the dual-mode sorption model to allow for a more realistic comparison with those obtained from transient permeation experiments. Surprisingly, $CO_2$ has a lower diffusion coefficient in MMM-1 and MMM-2 than in PIM-1, whereas $CH_4$ and $N_2$ have higher diffusion coefficients in MMM-1 and MMM-2. These disparities are due to competing effects of silica surfaces: the formation of larger pores, which improves gas diffusion, and a quadrupole–dipole interaction between gas molecules and silica surface hydroxyl groups, which retards gas diffusion. More specifically, the former significantly affects $CH_4$ and $N_2$ diffusivities because $N_2$ has a weak quadrupole moment and $CH_4$ has no quadrupole moment, whereas $CO_2$ diffusion is more susceptible to the latter due to the strong quadrupole–dipole interaction between $CO_2$ and surface hydroxyls. These findings provide intriguing insights into the behavior of gas diffusion in the immediate vicinity of PIM-1/silica interfaces, which are inaccessible to experiments.

We intend to investigate the effect of silica surface modification on gas adsorption and diffusion behaviors in PIM-1/silica interfacial systems in the future. It has been demonstrated experimentally that surface modification of silica nanoparticles leads to the formation of micro-and mesopores in a polymer phase, resulting in the enhancement of gas diffusivity.[29,100] However, its molecular-level mechanisms still remain elusive, and molecular simulations would provide valuable insights that complement experimental observations.

## ASSOCIATED CONTENT

**Supporting Information**

Structure factors of PIM-1 membranes; force-field parameters of the Morse-style potential used for constructing amorphous silica structure; structural characteristics of amorphous silica; CWCA force-field parameters for silica surfaces in MMMs; system sizes of model MMMs before/after 21-step equilibration; the relationship between chemical potential and pressure for the bulk gas phases of $CO_2$, $CH_4$, and $N_2$; and gas density distributions in MMM-2 obtained from GCMC simulations (PDF)

## AUTHOR INFORMATION

**Corresponding Author**

*E-mail: yyoshimoto@fel.t.u-tokyo.ac.jp (Y.Y.)

**Author Contributions**

[#]Y.Y. and Y.T. equally contributed to this work.

**Notes**

The authors declare no competing financial interest.

ACKNOWLEDGMENT

This work was partly supported by Research and Education Consortium for Innovation of Advanced Integrated Science (CIAiS) and JSPS KAKENHI Grant No. 20K14644.

Supporting Information

# Gas Adsorption and Diffusion Behaviors in Interfacial Systems Composed of a Polymer of Intrinsic Microporosity and Amorphous Silica: A Molecular Simulation Study

*Yuta Yoshimoto,*[,†,#] *Yuiko Tomita,*[†,#] *Kohei Sato,*[†] *Shiori Higashi,*[‡] *Masafumi Yamato,*[‡] *Shu Takagi,*[†] *Hiroyoshi Kawakami,*[‡] *and Ikuya Kinefuchi*[†]

[†]Department of Mechanical Engineering, The University of Tokyo, 7-3-1 Hongo, Bunkyo-ku, Tokyo 113-8656, Japan

[‡]Department of Applied Chemistry, Tokyo Metropolitan University, 1-1 Minami-osawa, Hachioji, Tokyo 192-0397, Japan

## S1. STRUCTURE FACTORS OF PIM-1 MEMBRANES

The structure factor $S(q)$ is related to a radial distribution function $g(r)$ as[1]

$$S(q) = 1 + 4\pi\rho \int_0^\infty r^2 [g(r) - 1] \frac{\sin qr}{qr} dr \qquad (S1)$$

where $r$ is the interatomic distance and $\rho$ is the number density. $S(q)$ can be measured by neutron or X-ray scattering experiments.

Figure S1 compares the simulated structure factors of PIM-1 membranes (# of chains = 1 and 8) with experimentally obtained counterparts.[2,3] The structure factors of 1-chain and 8-chain PIM-1 membranes are similar, showing good agreement with the experimental counterparts.

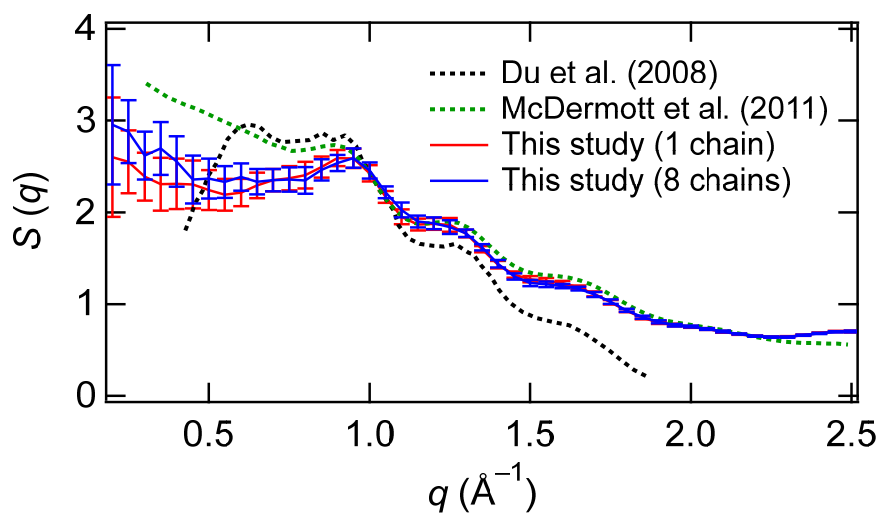

**Figure S1.** Comparison of the simulated structure factors of PIM-1 membranes (# of chains = 1 and 8) with experimental counterparts.[2,3] The simulated structure factors are averaged over five independent structures with error bars representing the standard deviations.

## S2. CONSTRUCTION OF AMORPHOUS SILICA

The interatomic interactions for bulk amorphous silica are represented by the Morse-style potential[4,5] as

$$U(r_{ij}) = \frac{q_i q_j}{4\pi\varepsilon_0} + D_0 \left\{ \exp\left[\gamma\left(1 - \frac{r_{ij}}{R_0}\right)\right] - 2\exp\left[\frac{\gamma}{2}\left(1 - \frac{r_{ij}}{R_0}\right)\right] \right\} \quad (S2)$$

where $r_{ij}$ is the interatomic distance between atom $i$ and $j$, $q_i$ is the charge of atom $i$, and $\varepsilon_0$ is the electric constant. The charges of Si and O atoms are 1.3e and −0.65e with e being the elementary charge, and the force field parameters are given in Table S1.

**Table S1.** Force Field Parameters of the Morse-Style Potential for Bulk Amorphous Silica

| Pair  | $R_0$ (Å) | $D_0$ (kcal/mol) | $\gamma$ |
|-------|-----------|------------------|----------|
| Si–Si | 3.7598    | 0.17733          | 15.3744  |
| O–O   | 3.791     | 0.5363           | 10.4112  |
| Si–O  | 1.628     | 45.997           | 8.6342   |

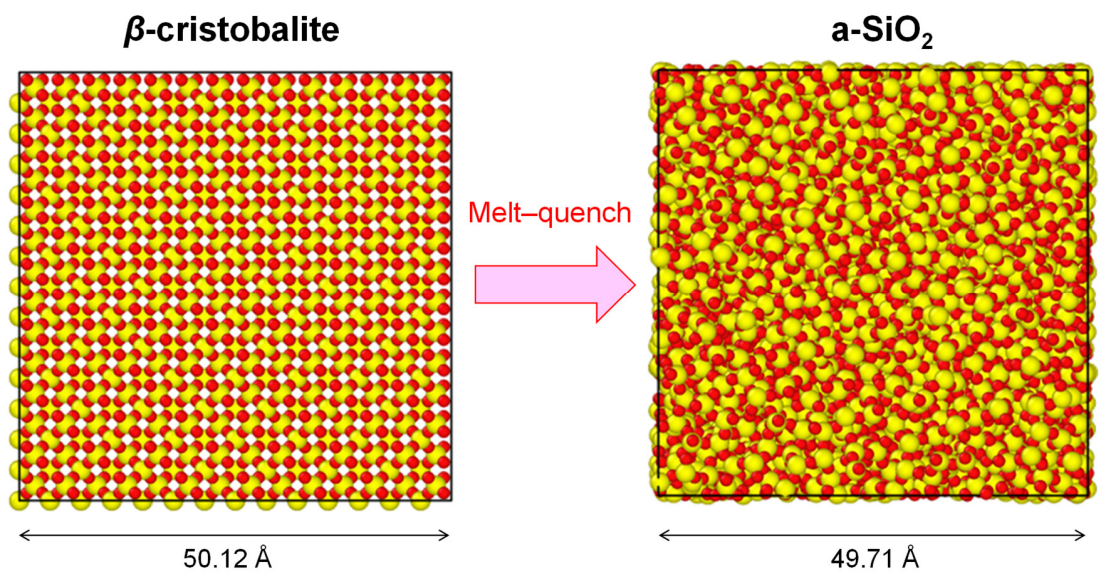

**Figure S2.** The *β*-cristobalite structure composed of 8232 atoms (left) and the amorphous silica structure at 300 K obtained by the melt–quench method (right). Silicon and oxygen atoms are represented by the yellow and red spheres, respectively.

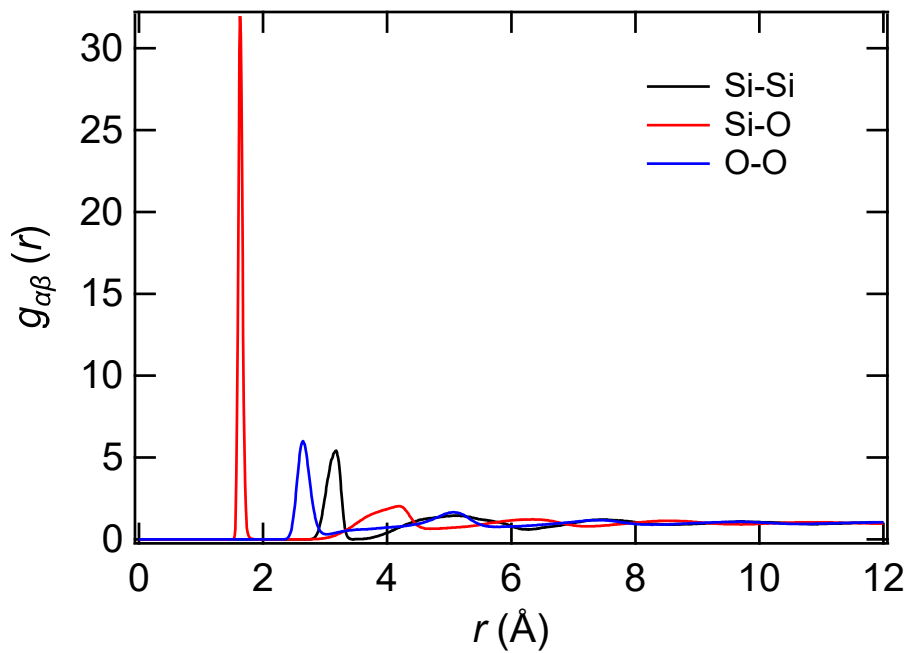

**Figure S3.** Partial radial distribution functions (PRDFs) for the bulk amorphous silica structure at 300 K. The positions of the minima after the first peaks of the Si–Si, Si–O, and O–O PRDFs are $R_{\text{Si-Si}} = 3.46$ Å, $R_{\text{Si-O}} = 2.03$ Å, and $R_{\text{O-O}} = 3.03$ Å, respectively. These values are used as cutoff radii to define coordination numbers (Tables S2 and S3).

**Table S2.** Average Coordination Numbers for the Bulk Amorphous Silica Structure at 300 K and 1 atm

| Si–Si | Si–O | O–Si | O–O |
|---|---|---|---|
| 4.02 | 4.00 | 2.00 | 6.17 |

**Table S3.** Fractions of Si Atoms with Coordination Number $Z_{Si-O} = 3, 4,$ and $5$, and Fractions of O Atoms with Coordination Number $Z_{O-Si} = 1, 2,$ and $3$ at 300 K and 1 atm

| $Z_{Si-O}$ | | | $Z_{O-Si}$ | | |
|---|---|---|---|---|---|
| 3 | 4 | 5 | 1 | 2 | 3 |
| 0.001 | 0.999 | 0.000 | 0.007 | 0.987 | 0.006 |

**Table S4.** Nonbonded Interaction Parameters for the CWCA Force Field[6,7]

| atom[a] | q | $\sigma$ (Å)[b] | $\epsilon$ (kcal/mol)[b] |
|---|---|---|---|
| Si | 0.9e | 3.8264 | 0.3 |
| $O_b$ | –0.45e | 3.118 | 0.1521 |
| $O_h$ | –0.66e | 3.1553 | 0.1521 |
| $H_h$ | 0.43e | 0.4 | 0.046 |

[a] $O_b$, $O_h$, and $H_h$ stand for bridging oxygens, hydroxide oxygens, and hydroxide hydrogens. [b] $\sigma$ and $\epsilon$ denote the Lennard-Jones (12-6) collision diameter and well depth, respectively.

**Table S5.** Bond-Stretching and Angle-Bending Parameters for the CWCA Force Field[7,8]

| Bond parameters[a] | | | Angle parameters[b] | | |
|---|---|---|---|---|---|
| bond | $K_b$ (kcal/(mol Å²)) | $R_0$ (Å) | angle | $K_a$ (kcal/(mol rad²)) | $\theta_0$ (deg) |
| Si–$O_h$ | 428.0 | 1.61 | Si–$O_h$–$H_h$ | 57.50 | 106.0 |
| $O_h$–$H_h$ | 545.0 | 0.96 | $O_h$–Si–$O_h$ | 89.62 | 116.26 |

[a] Bond-stretching potentials are given as $E_b = K_b(R - R_0)^2$. [b] Angle-bending potentials are given as $E_a = K_a(\theta - \theta_0)^2$.

## S3. CONSTRUCTION OF PIM-1/SILICA HYBRID SYSTEMS

**Table S6.** System Sizes of Five PIM-1 Membranes Composed of a Single Chain Before and After 21-step Equilibration for Use in the Construction of MMM-1

| System ID | Before 21-step equilibration | | | After 21-step equilibration | | |
|---|---|---|---|---|---|---|
| | $L_x$ (Å) | $L_y$ (Å) | $L_z$ (Å) | $L_x$ (Å) | $L_y$ (Å) | $L_z$ (Å) |
| 1 | 150 | 150 | 200 | 49.94 | 49.94 | 66.58 |
| 2 | 150 | 150 | 200 | 49.98 | 49.98 | 66.65 |
| 3 | 200 | 200 | 266 | 50.29 | 50.29 | 66.89 |
| 4 | 200 | 200 | 266 | 50.61 | 50.61 | 67.31 |
| 5 | 200 | 200 | 266 | 49.72 | 49.72 | 66.12 |

**Table S7.** System Sizes of Five PIM-1 Membranes Composed of Two Chains Before and After 21-Step Equilibration for use in the Construction of MMM-2

| System ID | Before 21-step equilibration | | | After 21-step equilibration | | |
|---|---|---|---|---|---|---|
| | $L_x$ (Å) | $L_y$ (Å) | $L_z$ (Å) | $L_x$ (Å) | $L_y$ (Å) | $L_z$ (Å) |
| 6 | 180 | 180 | 480 | 50.06 | 50.06 | 133.50 |
| 7 | 180 | 180 | 480 | 49.42 | 49.42 | 131.79 |
| 8 | 180 | 180 | 480 | 49.67 | 49.67 | 132.45 |
| 9 | 180 | 180 | 480 | 49.98 | 49.98 | 133.28 |
| 10 | 250 | 250 | 665 | 49.82 | 49.82 | 132.52 |

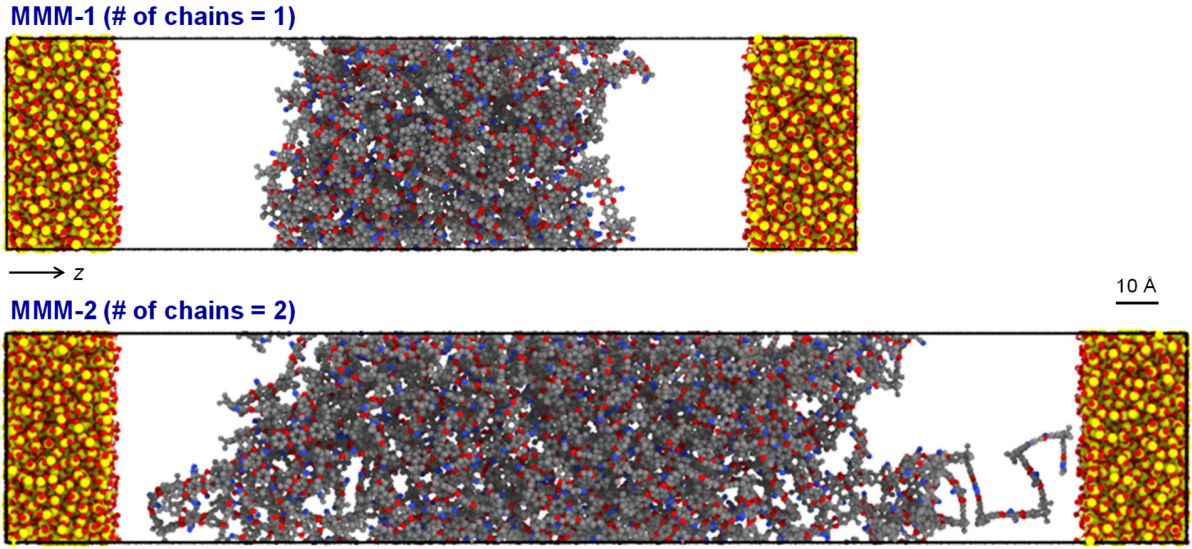

**Figure S4.** MMMs made of PIM-1 chains and amophous silica before 21-step equilibration simulations. The number of PIM-1 chains is one (MMM-1) and two (MMM-2). Carbon, oxygen, nitrogen, silicon, and hydrogen atoms are represented by the gray, red, blue, yellow, and white spheres, respectively.

## S4. RELATIONSHIP BETWEEN CHEMICAL POTENTIAL AND PRESSURE

The chemical potential can be expressed as follows:

$$\mu = \mu^{\text{id}} + \mu^{\text{ex}} \tag{S3}$$

where $\mu^{\text{id}}$ is the ideal gas contribution and $\mu^{\text{ex}}$ is the excess chemical potential due to energetic interactions. The former is given as

$$\mu^{\text{id}} = k_\text{B} T \ln \frac{p \Lambda^3}{k_\text{B} T} \tag{S4}$$

where $k_\text{B}$ is the Boltzmann constant, $\Lambda$ is the thermal de Broglie wavelength, $p$ is the pressure, and $T$ is the temperature.

To account for the nonideality of the gases at high pressure, bulk gas-phase GCMC simulations at 300 K in the µVT ensemble are performed to obtain the relationship between the chemical potential and pressure for $CO_2$, $CH_4$, and $N_2$. In each GCMC simulation, the chemical potential ($\mu$) is specified as an input parameter using eq S4 with $p$ being the target pressure (1, 2, 5, 10, 20, 30, and 40 atm); in reality, a final pressure obtained from the GCMC simulation is different from the target pressure due to the effect of $\mu^{ex}$. To ensure reliable statistics for each gas species, the simulation box size is adjusted for each $\mu$ such that the system contains at least 100 molecules after the GCMC simulation has sufficiently converged.

Figure S5 depicts the obtained relationship between the pressure and chemical potential. $CH_4$ and $N_2$ reasonably follow eq S4 up to a pressure of ~ 40 atm, showing only slight differences between $\mu$ and $\mu^{id}$ (i.e., small $\mu^{ex}$). Meanwhile, $CO_2$ begins to deviate from eq S4 with increasing pressure (> 20 atm), showing a nonnegligible effect of $\mu^{ex}$. Table S8 summarizes the relationship between the chemical potential and pressure for each gas species.

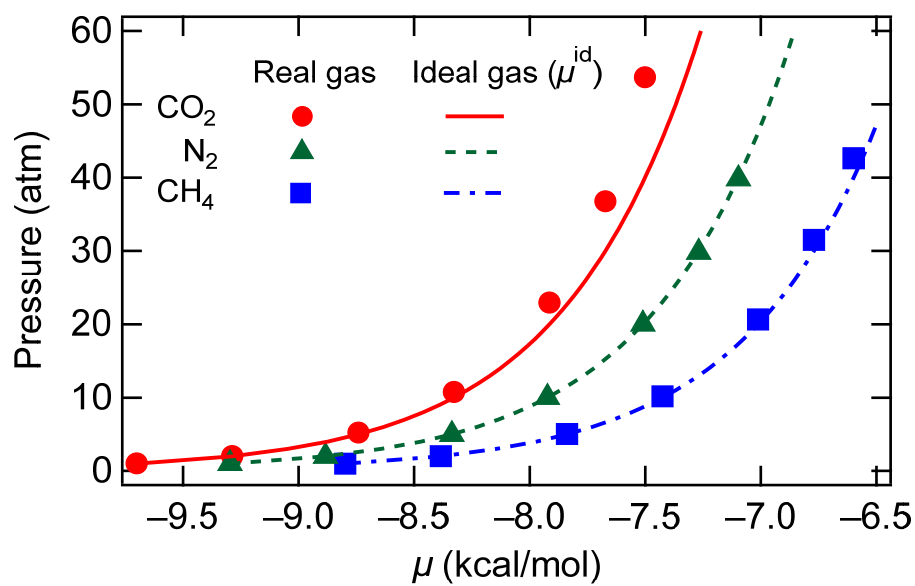

**Figure S5.** Relationship between the pressure and chemical potential for $CO_2$, $CH_4$, and $N_2$. The solid, dashed-dotted, and dashed lines represent ideal gas relations for $CO_2$, $CH_4$, and $N_2$, respectively, calculated using eq S4.

**Table S8.** Relationship between the Pressure and Chemical Potential for $CO_2$, $CH_4$, and $N_2$.

| Gas | $\mu$ (kcal/mol) | $p$ (atm) |
| --- | --- | --- |
| $CO_2$ | −9.700 | 1 |
|  | −9.287 | 2 |
|  | −8.741 | 5 |
|  | −8.328 | 11 |
|  | −7.914 | 23 |
|  | −7.673 | 37 |
|  | −7.501 | 54 |
| $CH_4$ | −8.798 | 1 |
|  | −8.385 | 2 |
|  | −7.838 | 5 |
|  | −7.425 | 10 |
|  | −7.012 | 21 |
|  | −6.770 | 32 |
|  | −6.599 | 43 |
| $N_2$ | −9.296 | 1 |
|  | −8.883 | 2 |
|  | −8.337 | 5 |
|  | −7.924 | 10 |
|  | −7.510 | 20 |
|  | −7.269 | 30 |
|  | −7.097 | 40 |

## S5. GAS DENSITY DISTRIBUTIONS IN MMM-2

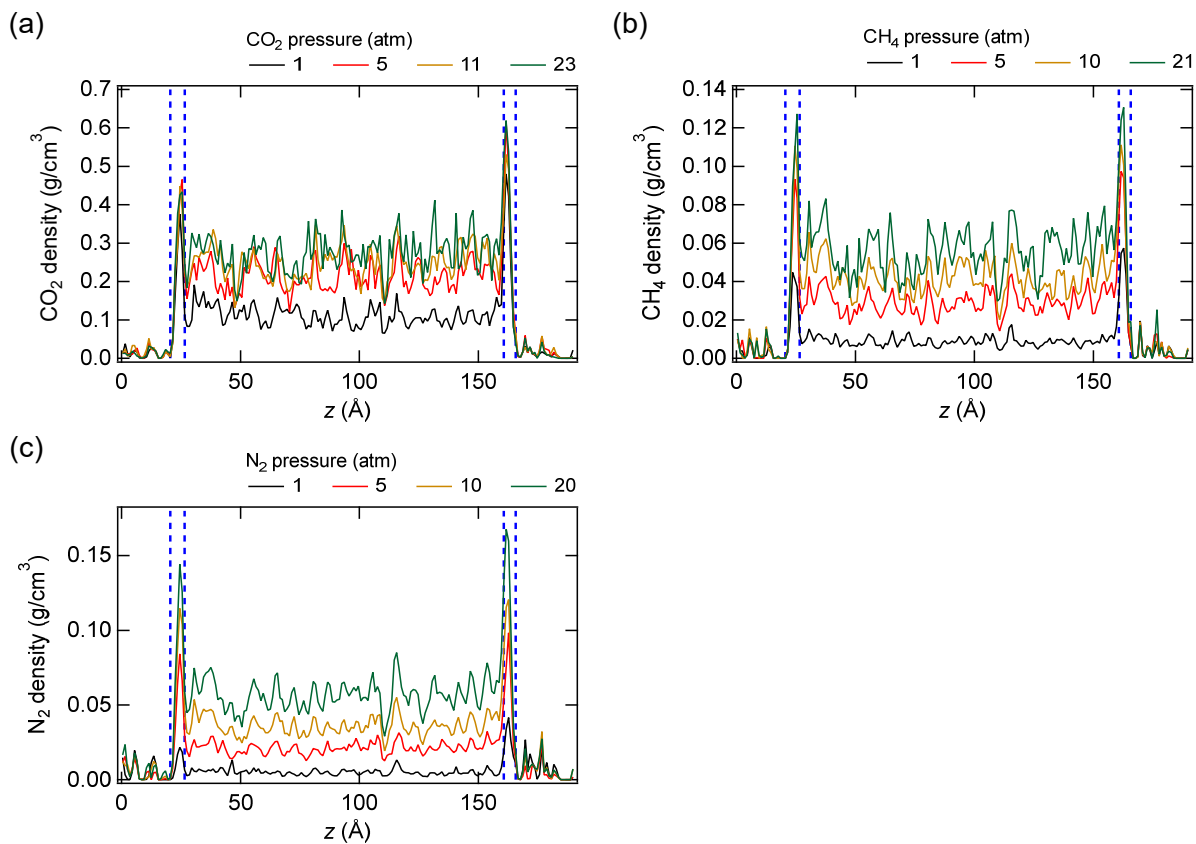

**Figure S6.** Gas density distributions in representative MMM-2; $CO_2$ at pressures ranging from 1 to 23 atm (a), $CH_4$ at pressures ranging from 1 to 21 atm (b), and $N_2$ at pressures ranging from 1 to 20 atm (c). According to Figure 2b, the blue dashed lines represent PIM-1/silica mixing regions.